\documentclass[11pt,a4paper,reqno]{amsart}
\usepackage{amsthm,amsmath,amsfonts,amssymb,amsxtra,appendix,bookmark,dsfont,latexsym,bm,hyperref,delarray,color,euscript,amsgen,amsbsy,amsopn,amscd,latexsym,mathrsfs}

\makeatletter

\usepackage{hyperref}

\makeatletter
\def\@tocline#1#2#3#4#5#6#7{\relax
  \ifnum #1>\c@tocdepth 
  \else
    \par \addpenalty\@secpenalty\addvspace{#2}%
    \begingroup \hyphenpenalty\@M
    \@ifempty{#4}{%
      \@tempdima\csname r@tocindent\number#1\endcsname\relax
    }{%
      \@tempdima#4\relax
    }%
    \parindent\z@ \leftskip#3\relax \advance\leftskip\@tempdima\relax
    \rightskip\@pnumwidth plus4em \parfillskip-\@pnumwidth
    #5\leavevmode\hskip-\@tempdima
      \ifcase #1
       \or\or \hskip 1em \or \hskip 2em \else \hskip 3em \fi%
      #6\nobreak\relax
      \dotfill
      \hbox to\@pnumwidth{\@tocpagenum{#7}}
    \par
    \nobreak
    \endgroup
  \fi}
\makeatother

\newtheorem{theorem}{Theorem}
\newtheorem{lemma}[theorem]{Lemma}

\newtheorem{conjecture}[theorem]{Conjecture}
\theoremstyle{definition}

\theoremstyle{remark}

\newcommand{\ii}{\infty}
\newcommand\R{{\ensuremath {\mathbb R} }}
\newcommand\C{{\ensuremath {\mathbb C} }}
\newcommand\N{{\ensuremath {\mathbb N} }}

\newcommand\1{{\ensuremath {\mathds 1} }}

\renewcommand{\>}{\rangle}
\newcommand\nn{\nonumber}

\renewcommand\phi{\varphi}

\newcommand{\bH}{\mathbb{H}}

\newcommand{\bW}{\mathbb{W}}

\newcommand{\gH}{\mathfrak{H}}

\newcommand{\gS}{\mathfrak{S}}

\newcommand{\cF}{\mathcal{F}}
\newcommand{\cN}{\mathcal{N}}
\newcommand{\cD}{\mathcal{D}}
\newcommand{\cH}{\mathcal{H}}

\newcommand{\cZ}{\mathcal{Z}}

\renewcommand{\epsilon}{\varepsilon}
\newcommand\pscal[1]{{\ensuremath{\left\langle #1 \right\rangle}}}
\newcommand{\norm}[1]{ \left| \! \left| #1 \right| \! \right| }
\DeclareMathOperator{\tr}{{\rm Tr}}
\DeclareMathOperator{\Tr}{{\rm Tr}}

\renewcommand{\ge}{\geqslant}
\renewcommand{\le}{\leqslant}
\renewcommand{\geq}{\geqslant}
\renewcommand{\leq}{\leqslant}
\renewcommand{\hat}{\widehat}

\newcommand{\cHcl}{\cH_{\rm cl}}

\newcommand{\lam}{\lambda}
\newcommand{\eps}{\varepsilon}

\newcommand{\Varq}{\mathrm{Var}^{(0)}}
\newcommand{\Vars}{\mathrm{Var}^{(s)}}

\newcommand{\av}[1]{\left\langle#1\right\rangle_{\mu_0}}

\begin{document}

\title[Derivation of renormalized Gibbs measures]{Derivation of renormalized Gibbs measures from equilibrium many-body quantum Bose gases}

\author[M. Lewin]{Mathieu LEWIN}
\address{CNRS \& Universit\'e Paris-Dauphine, PSL University, CEREMADE, Place de Lattre de Tassigny, F-75016 PARIS, France} 
\email{mathieu.lewin@math.cnrs.fr}

\author[P.T. Nam]{Phan Th\`anh NAM}
\address{Department of Mathematics, LMU Munich, Theresienstrasse 39, 80333 Munich, Germany} 
\email{nam@math.lmu.de}

\author[N. Rougerie]{Nicolas ROUGERIE}
\address{Universit\'e Grenoble-Alpes \& CNRS,  LPMMC (UMR 5493), B.P. 166, F-38042 Grenoble, France}
\email{nicolas.rougerie@grenoble.cnrs.fr}

\date{May 28, 2019}

\begin{abstract} We review our recent result on the rigorous derivation of the renormalized Gibbs measure from the many-body Gibbs state in 1D and 2D. The many-body renormalization is accomplished by simply tuning the chemical potential in the grand-canonical ensemble, which is analogous to the Wick ordering in the classical field theory. 
\end{abstract}

\maketitle

\tableofcontents

\maketitle

\section{Introduction}

A major challenge in mathematical physics is to understand the Bose-Einstein phase transition from the first principles of quantum mechanics. It is expected that if the temperature is sufficiently low, then a macroscopic fraction of particles occupies a common quantum state. This phenomenon has been rigorously justified in several works in the last decades; we refer to \cite{LieSeiSolYng-05,Schlein-08,Golse-13,Rougerie-14,Lewin-15,BenPorSch-15} for reviews. On the other hand, a superposition state of condensates is expected to form just above the critical temperature. Some recent results in this direction will be discussed below. 

We consider a Bose gas in the torus $\Omega=\mathbb{T}^d$ described by the grand-canonical Hamiltonian
$$
\mathbb{H}_\lambda = \int_{\Omega} a_x^* h_x a_x d x + \frac{\lambda}{2} \iint_{\Omega \times \Omega}  a_x^* a_y^* w(x-y) a_x a_y dx dy 
$$
on the bosonic Fock space 
$$
\mathfrak{F} = {\mathbb C} \oplus L^2(\Omega)\oplus \ldots \oplus L^2_{\rm sym}(\Omega^n)  \oplus \ldots 
$$
Here $h=-\Delta-\nu$ with $-\Delta$ the usual kinetic energy operator (Laplacian with periodic boundary condition) and $\nu \in {\mathbb R}$ a chemical potential ensuring $h>0$; $w \ge 0$ is a periodic interaction potential and $\lambda\ge 0$ represents the interaction strength; and $a^\dagger_x,a_x$ are the usual creation/annihilation operators which satisfy the canonical commutation relations 
$$
 \left[a_x, a_y\right] = 0 = [a^\dagger_x, a ^*_y] = 0, \quad [a_x, a ^*_y] = \delta (x-y).
$$

We are interested in the equilibrium (Gibbs) state at a positive temperature $T>0$
\begin{equation} \label{eq:intro quantum Gibbs}
\Gamma_{\lambda} = \mathcal{Z}_\lambda^{-1} \exp\left( - \mathbb{H}_{\lambda}/T \right) , \quad \mathcal{Z}_\lambda = {\rm Tr} \left[ \exp\left( - \mathbb{H}_{\lambda}/T \right) \right].
\end{equation}

\subsection{Free gas} In the non-interacting case $\lambda=0$, the free Gibbs state $
\Gamma_{0}$ is a quasi-free state and many of its properties can be analyzed explicitly using  Wick's theorem. In particular, its $k$-body density matrix, defined as a trace class operator on $L^2_{\rm sym}(\Omega^k)$ with kernel 
$$
\Gamma^{(k)}(x_1,...,x_k; y_1,...,y_k)= (k!)^{-1} {\rm Tr}\left[ a^\dagger_{x_1}... a^\dagger_{x_k} a_{y_1}... a_{y_k} \Gamma \right], 
$$
is  
\begin{align} \label{eq:com-G0k}
\Gamma_{0}^{(k)}=\left(\frac{1}{e^{h/T}-1}\right)^{\otimes k} \underset{T\to \infty}{\approx} \quad T^k \,(h^{-1})^{\otimes k}.
\end{align}
Here we used the convention that $A ^{\otimes k}$ denotes the projection of the tensor product on the symmetric subspace (the full density matrix includes the exchange terms).

The  limit can be rewritten in a fancy way \cite[Lemma 3.3]{LewNamRou-14d}
\begin{align}\label{eq:factor-form-free}
(h^{-1})^{\otimes k}=\frac{1}{k!}\int  |u^{\otimes k}\rangle\langle u^{\otimes k}|\,d\mu_0(u)
\end{align}
using the infinite-dimensional Gaussian measure 
$$d\mu_0(u)=``z_0^{-1}e^{-\langle u, hu\rangle}\,du "= \bigotimes_{j=1}^\infty \left( \frac{\lambda_j}{\pi}e^{-\lambda_j|\alpha_j|^2}\,d\alpha_j\right), \quad \alpha_j=\langle u_j, u\rangle.$$
Here we have used the spectral decomposition 
$$h=\sum_{j=1}^\infty \lambda_j |u_j\rangle \langle u_j|$$ 
and $\mu_0$ can be defined properly by its cylindrical projections on finite dimensional eigenspaces $\1(h\le K)$ as in  \cite[Lemma 1]{Skorokhod-74}. It turns out that $\mu_0$ is well defined on the (possibly negative) Sobolev space 
\begin{align} \label{eq:def-H1-p}
\mathfrak{H}^{1-p}=\left\{u=\sum_{j=1}^\infty \alpha_j u_j \,:\, \sum_{j=1}^\infty \lambda_j^{1-p} |\alpha_j|^2 <\infty \right\}
\end{align}
provided that $\Tr [h^{-p}]<\infty$ for some $p \ge 1$. On the other hand, $\mu_0$ is supported completely outside $\mathfrak{H}^{1-q}$ if ${\rm Tr}[h^{-q}]=\infty$. This so-called zero-one law follows from Fernique's theorem; see \cite[Section 3.1]{LewNamRou-14d} for further details. 

Note that the full operator $\Gamma_{0}^{(k)}$ can be also written in the form \eqref{eq:factor-form-free} by a $T$-dependent measure where one replaces $h^{-1}$ by $(e ^{h/T} - 1)^{-1}$ as covariance. This measure is supported on regular functions and the domain issue only emerges  in the limit.

One way to interpret the emergence of the Gaussian measure $\mu_0$ in the convergence
\begin{align} \label{eq:CV-DM-free}
\frac{k!}{T^k} \Gamma_{0}^{(k)} \underset{T\to \infty}{\longrightarrow} \int  |u^{\otimes k}\rangle\langle u^{\otimes k}|\,d\mu_0(u)
\end{align}
is as follows. We write  $$
\frac{\bH_0}{T}=\frac{1}{T}\int_{\Omega} a_x^* h_x a_x d x = \int_{\Omega} b_x^* h_x b_x d x
$$
where the new operators $b_x=a_x/\sqrt{T}$ almost commute in the large $T$ limit. Formally replacing operators $b_x,b_x^*$ by functions $u(x), \overline{u(x)}$ leads to the quantum-classical correspondence  
\begin{align} \label{eq:semiclassical-heu}
\boxed{\Gamma_0=\mathcal{Z}_0^{-1}e^{-\bH_0/T} \quad \longleftrightarrow \quad d\mu_0 = ``z_0^{-1}e^{-\langle u, hu\rangle}\,du ".}
\end{align}
Thus \eqref{eq:CV-DM-free} is a rigorous justification of the semiclassical approximation \eqref{eq:semiclassical-heu}.

\subsection{Interacting gas} Now we turn on the interaction and focus on the mean-field regime 
$$ T\to \infty, \quad \lambda = T^{-1}.$$
The specific choice of $\lambda$ allows us to write  
$$
\frac{\mathbb{H}_\lambda}{T} =\int_{\Omega} b^\dagger_x (-\Delta_x-\nu) b_x \, d x+ \frac{1}{2} \iint_{\Omega \times \Omega} b^\dagger_x b^\dagger_y w (x-y) b_x b_y \,dx\,dy 
$$
with $b_x=a_x/\sqrt{T}$ and obtain the formal analogue of \eqref{eq:semiclassical-heu} 
\begin{align} \label{eq:semiclassical-heu-int}
\boxed{\Gamma_\lambda=\mathcal{Z}_\lambda^{-1}e^{-\bH_\lambda/T} \quad \longleftrightarrow \quad d\mu = ``z^{-1}e^{-\langle u, hu\rangle - \mathcal{D}[u]}\,du " = z_{r}^{-1} e^{-\mathcal{D}[u]} d\mu_0(u).}
\end{align}

The so-obtained $\mu$ (with an appropriate nonlinear functional $\mathcal{D}[u]$) is called a {\em nonlinear Gibbs measure}. This  measure played a central role in constructive quantum field theory in the 1970s (see ~\cite{Simon-74,GliJaf-87,DerGer-13} for reviews). Since the quantum Gibbs state is obviously invariant under the many-body Schr\"odinger flow, the correspondence \eqref{eq:semiclassical-heu-int} strongly confirms the fact that the nonlinear Gibbs measure in \eqref{eq:semiclassical-heu-int} is a natural candidate for an invariant measure under the NLS flow
$$
i\partial_t u = (-\Delta -\nu+ w*|u|^2) u
$$
(possibly under an appropriate renormalization). It is the latter property that makes the Gibbs measure (and its variants) very useful in many studies on nonlinear dispersive equations  \cite{LebRosSpe-88,Bourgain-96,BurThoTzv-09} and nonlinear stochastic PDEs \cite{PraDeb-03,LorGub-09,Hairer-14}. Thus a rigorous justification for \eqref{eq:semiclassical-heu-int}, e.g. in an analogue of \eqref{eq:CV-DM-free}, is desirable and it is the goal of our papers  \cite{LewNamRou-14d,LewNamRou-17,LewNamRou-18}.

In one dimension,  $h^{-1}$ is trace class and  $\mu_0$ is supported on $L^2$-functions. Therefore, we can simply take 
\begin{align} \label{eq:def-Iu}
\mathcal{D}[u]=\frac{1}{2}\iint_{\Omega \times \Omega} |u(x)|^2\, w(x-y)|u(y)|^2 \,dx\,dy,  
\end{align} 
which is nonnegative (as $w\ge 0$) and finite $\mu_0$-almost surely, e.g. for $w\in L^1+L^\infty$. Thus $\mu$ in \eqref{eq:semiclassical-heu-int} is a probability measure on the same support of $\mu_0$. In \cite[Theorem 5.3]{LewNamRou-14d} we proved the following 
\begin{theorem}[Emergence of 1D Gibbs measure] \label{thm:1D} Let $d=1$ and let $w=w_1+w_2$ with $0\le w_1\in L^\infty(\R)$ and $w_2$ a positive measure with finite mass. Then $\mu$ in \eqref{eq:semiclassical-heu-int}-\eqref{eq:def-Iu} is well-defined. Moreover, for any $\nu\in \mathbb{R}$ such that $h>0$, in the limit $\lambda = T^{-1}\to 0$ we have 
\begin{align} \label{eq:CV-DM-int-1D}
\Tr \left| \frac{k!}{T^k}\Gamma_{\lambda}^{(k)} - \int_\mathfrak{H} |u^{\otimes k}\rangle\langle u^{\otimes k}|\,d\mu(u) \right| \to 0, \quad \forall k\ge 1.
\end{align} 
\end{theorem}

This result was recovered later by Fr\"ohlich-Knowles-Schlein-Sohinger  \cite{FroKnoSchSoh-16} using a different method (see also \cite{FroKnoSchSoh-17} for an investigation of the time-dependent problem).  A finite-dimensional version of \eqref{eq:CV-DM-int-1D} was proved earlier by Gottlieb \cite{Gottlieb-05} (see also \cite{Knowles-09,Rougerie-14}). The proof of \eqref{eq:CV-DM-int-1D} below requires infinite dimensional semiclassical analysis.  

In higher dimensions, $h^{-1}$ is no longer trace class and $\mu_0$ is supported on negative Sobolev spaces. Therefore, the naive choice  \eqref{eq:def-Iu} does not work as the corresponding functional is infinite $\mu_0$-almost surely (for all nontrivial smooth function $w$). Thus some sort of renormalization is needed. 

It turns out that if $w$ is a nice potential, then a  Wick ordering is sufficient. The idea, going back to Nelson~\cite{Nelson-66}, amounts to remove a uniform, infinite constant from the mass $\int |u|^2$, leading to the renormalized interaction energy
\begin{align} \label{eq:def-Iu-ren}
\mathcal{D}[u]=  \frac{1}{2} \iint_{\Omega^2} \left( |u(x)| ^2 - \left\langle |u(x)| ^2 \right\rangle_{\mu_0} \right) w (x-y) \left( |u(y)| ^2 - \left\langle |u(y)| ^2 \right\rangle_{\mu_0} \right) dx dy
\end{align}
where $\langle \cdot \rangle_{\mu_0}$ denotes the expectation in the free Gibbs measure $\mu_0$. In fact, \eqref{eq:def-Iu-ren} is not the full Wick ordering used by Nelson as the case we deal with here is much simpler than Wick ordering $|u|^4$. 

The formula \eqref{eq:def-Iu-ren} should be interpreted properly using cylindrical projections on finite dimensional eigenspaces of $h$ and taking a limit (as in the definition of $\mu_0$). We will assume that the interaction $w$ is of positive type, namely its Fourier transform is positive
$$
\widehat w(k) = \int_{\Omega} w(x) e^{-i k\cdot x} dx \ge 0,
$$
which ensures that the renormalized interaction is positive. We will also restrict to $d \le 3$ which ensures the important condition $\tr [h ^{-2}] < \infty$ for the analysis. To be precise, we have (see \cite[Lemma 5.3]{LewNamRou-18})

\begin{lemma}[Renormalized Gibbs measure]\label{lem:re-interaction} Let $d\le 3$, $h=-\Delta+\kappa>0$ and $0\le \widehat w \in \ell^1(2\pi \mathbb{Z}^d)$. Then $\cD[u]$ in \eqref{eq:def-Iu-ren} can be defined as the strong limit in $L^1(d\mu_0)$ of the sequence
$$
 \mathcal{D}_K[u]:=\frac{1}{2}\iint_{\Omega^2} \left(|P_K u(x)|^2-\av{|P_K u(x)|^2}\right) w(x-y)  \left(|P_Ku(y)|^2-\av{|P_Ku(y)|^2}\right)\,dx\,dy
$$
with $P_K=\1(h\le K)$. Consequently, $d\mu(u)$ in \eqref{eq:semiclassical-heu-int} is well-defined as a probability measure. 
\end{lemma}

Now let us turn to the quantum model. We consider the free Gibbs state associated with $h=-\Delta+\kappa>0$ and its density 
$$N_0=\langle a_x^* a_x \rangle_{\Gamma_0}= \sum_{k\in (2\pi {\mathbb Z})^d}  \frac{1}{e^{\frac{|k|^2+\kappa}{T}} -1}.$$
(This is independent of $x$ due to the translation invariance.) The renormalized quantum  interaction is
\begin{align} \label{eq:def-W}
\lambda \bW& = \frac{\lambda}{2}\iint_{\Omega^2} (a_x^* a_x - N_0) w(x-y)  ( a_y^*a_y -N_0) \, dx dy \\
&= \frac{ \lambda}{2} \iint_{\Omega^2}   a_x^* a_y^* w(x-y) a_x a_y dx dy +  \left(\lambda \frac{w(0)}{2} - \lambda \hat w(0)N_0\right) \int_\Omega a_x^* a_x dx +  \frac{\lambda}{2}N_0^2 \widehat w(0). \nn
\end{align}
This leads to the adjusted chemical potential 
\begin{align}\label{eq:def-nu-ren}
\boxed{\nu =  \lambda \hat w(0)N_0 -\kappa}
\end{align}
which is proportional to $\log T$ if $d=2$ and $\sqrt{T}$ if $d=3$. Here the factor $\lambda w(0)/2\sim O(T^{-1})$ is negligible, while the energy shift $\lambda N_0^2/2$ is huge but irrelevant to the Gibbs state. 

With the choice of $\nu$ in \eqref{eq:def-nu-ren}, we expect that the quantum-classical correspondence \eqref{eq:semiclassical-heu-int} remains valid. The precise statements will be discussed in the next section. 

\section{A conjecture and rigorous results}

From the previous discussion, it is natural to suggest the following

\begin{conjecture}[Emergence of renormalized Gibbs measure]  Let $d=2,3$, $h=-\Delta+\kappa>0$ and $0\le \widehat w \in \ell^1(2\pi \mathbb{Z}^d)$. Let $\nu$ as in \eqref{eq:def-nu-ren} and consider the Gibbs state $\Gamma_\lambda=\mathcal{Z}_\lambda e^{-\bH_\lambda/T}$ with 
$$
\mathbb{H}_\lambda = \int_{\Omega} a_x^* (-\Delta-\nu) a_x d x + \frac{\lambda}{2} \iint_{\Omega \times \Omega}  a_x^* a_y^* w(x-y) a_x a_y dx dy. 
$$
Then in the limit $\lambda=T^{-1}\to 0$ we obtain the renormalized Gibbs measure: 
\begin{align} \label{eq:cv-Gibbs-state-measure}
\Tr \left| \frac{k!}{T^k} \Gamma_\lambda^{(k)} - \int |u^{\otimes k} \rangle \langle u^{\otimes k} | d\mu(u) \right|^p \to 0, \quad \forall k\ge 1, \forall p>d/2.
\end{align}
\end{conjecture}


The first attempt to resolve the above conjecture is due to  Fr\"ohlich-Knowles-Schlein-Sohinger \cite{FroKnoSchSoh-16}.  They proved \eqref{eq:cv-Gibbs-state-measure} for $d=2,3$ and $p=2$, but with the modified quantum state
\begin{equation}\label{eq:Fro et al}
\Gamma_{\lambda} ^\eta = \frac{1}{\mathcal{Z}_\lambda ^\eta} \exp\left(- \frac{\eta}{2T} \mathbb{H}_0 \right) \exp\left(- \frac{\mathbb{H}_\lambda- \eta \mathbb{H}_0 }{T}\right) \exp\left(- \frac{\eta}{2T} \mathbb{H}_0 \right)
\end{equation}
for a fixed parameter $0 < \eta < 1$, using a direct analysis of the reduced density matrices and Borel summation method for divergent series. It is unclear whether this approach can be used to treat the true quantum Gibbs state. 

Very recently, we were able to prove the above conjecture in two dimensions \cite{LewNamRou-18}.  The three dimensional case remains open.  Our result reads as follows.

\begin{theorem}[Emergence of 2D Gibbs measure] \label{thm:main}If $d=2$ and if $0\le \hat w\in \ell^1(2\pi \mathbb{Z}^d, (1+|k|^\alpha))$ for some $\alpha>0$, then in the limit $\lambda=T^{-1} \to 0$ we have 
$$
\Tr \left| \frac{k!}{T^k} \Gamma_\lambda^{(k)} - \int |u^{\otimes k} \rangle \langle u^{\otimes k} | d\mu(u) \right|^p \to 0, \quad \forall k\ge 1, \forall p>1,
$$
and 
$$
\Tr \left| \frac{1}{T} \left( \Gamma_\lambda^{(1)} -  \Gamma_0^{(1)} \right) - \int |u \rangle \langle u | \Big( d\mu(u) - d\mu_0(u) \Big) \right| \to 0.
$$
\end{theorem}

The analysis in  \cite{FroKnoSchSoh-16} and \cite{LewNamRou-18} also covers the inhomogeneous case, when one considers bosons in $\R^d$ with an external trapping potential $V(x)$. In this case, the renormalization of the quantum problem becomes more complicated because the density of the free Gibbs state depends on the position $x$, and it turns out that the limiting Gibbs measure will be associated with a potential $V_\infty$ different from $V$ ($V_\infty$ is determined from $V$ and $w$ via a counter-term problem). We refer to the original papers for further details in this direction.  

Below we will explain the main ingredients of the proof of Theorem \ref{thm:main}.

\section{Ingredients of the proof}

\subsection{Variational approach.} Our general strategy is based on Gibbs' variational principle that $\Gamma_\lambda$ is the unique mininizer of the free energy 
$$
-\log \mathcal{Z}_\lambda  = \inf \left\{ T^{-1}\Tr [\bH_\lambda \Gamma] +  \Tr [\Gamma \log \Gamma] : 0\le \Gamma \le 1, \Tr [\Gamma]=1 \right\}.
$$
In the following let us shift $\bH_\lambda$ by the constant $\lambda \widehat w(0) N_0^2/2$ (which does not change the Gibbs state), so that we can write 
$$\bH_\lambda = \bH_0 + \lambda \bW$$
with $\bW$ the renormalized interaction as in \eqref{eq:def-W}. Using the free Gibbs state $\Gamma_0$ as a reference state, we find that $\Gamma_{\lambda}$ is the unique minimizer for the {relative free energy}
\begin{align} \label{eq:rel-free}
\boxed{-\log \frac{\mathcal{Z}_\lambda}{\mathcal{Z}_0}= \inf \left\{  \mathcal{H}(\Gamma,\Gamma_{0}) + T^{-2} {\rm Tr} [\bW \Gamma] : 0\le \Gamma \le 1, \Tr [\Gamma]=1 \right\}}
\end{align}
with the quantum {relative entropy} 
$$\mathcal{H}(\Gamma,\Gamma')= {\rm Tr}_{\mathfrak{F}}\big(\Gamma(\log\Gamma-\log\Gamma')\big) \ge 0.$$

Similarly, the Gibbs measure $\mu$ is the unique minimizer of 
\begin{align} \label{eq:rel-free-classical}
\boxed{-\log z_r = \inf \left\{ \mathcal{H}_{\rm cl}(\eta, \mu_0) +\int \mathcal{D}[u]  \,d\eta(u) : \eta \text{ a probability measure} \right\}}
\end{align} 
with the classical relative entropy 
$$ \mathcal{H}_{\rm cl} (\eta,\eta'):= \int \frac{d\eta}{d\eta'}(u)\log\left(\frac{d\eta}{d\eta'}(u)\right)\,d\eta'(u) \ge 0.$$

Our goal is to relate the quantum problem \eqref{eq:rel-free} and its classical analogue \eqref{eq:rel-free-classical}. 

\subsection{De Finetti measure} Let us recall a simple variant of the quantum de Finetti Theorem in Fock space \cite[Theorem~4.2]{LewNamRou-14d} (whose proof goes back to the analysis of~\cite{AmmNie-08,LewNamRou-14}).

\begin{theorem}[De Finetti measure] Let  $\{\Gamma_n\}$ be a sequence of quantum states and let $T_n\to +\infty$ such that for some $p\ge 1$,
\begin{equation} \label{eq:a-priori0Tn}
\limsup_{T_n\to \infty} {\rm Tr} \left| \frac{\Gamma_n^{(k)}} {T_n^k} \right|^p < \infty, \quad \forall k\ge 1.
\end{equation}
Then for any one-body self-adjoint operator $h>0$ satisfying $\Tr[h^{-p}]<\infty$, there exist a subsequence of $\{\Gamma_n\}$ and a probability measure $\eta$ supported on the Sobolev-type space $\mathfrak{H}^{1-p}$ defined from $h$ similarly to \eqref{eq:def-H1-p} such that 
$$
\frac{k!}{T_n^k} \Gamma_{n}^{(k)} \rightharpoonup \int_{\mathfrak{H}^{1-p}} |u^{\otimes k}\rangle \langle u^{\otimes k}| \,d\eta(u), \quad \forall k\ge 1
$$
weakly-$\ast$ in the Schatten space $\mathfrak{S} ^p$. The measure $\eta$ is called the de Finetti measure  (or Wigner measure) of the quantum states $\{\Gamma_{n}\}$ at scale $T_n^{-1}$. 
\end{theorem}

In applications, we can verify \eqref{eq:a-priori0Tn} for the interacting Gibbs state $\Gamma_\lambda$ in the Hilbert-Schmidt topology $p=2$. In fact, from the positivity of the heat kernel and the interaction, a standard argument using the Trotter product formula leads to the kernel estimate \cite[Lemma 4.3]{LewNamRou-17}
$$
0\le \Gamma_\lambda^{(k)}(X_k;Y_k)\le C_k\Gamma_{0}^{(k)}(X_k;Y_k). 
$$
Combining with the formula for the free density matrices \eqref{eq:com-G0k} leads to
$$ \Tr \left| \frac{\Gamma_\lambda^{(k)} }{T^k} \right|^2 \le C_k  \left| \frac{\Gamma_0^{(k)} }{T^k} \right|^2\le \widetilde{C}_k, \quad \forall k\ge 1$$
where the constant $\widetilde{C}_k$ is uniform in $T$ and depends on $h$ only via $\Tr[h^{-2}]$. 

In one dimension (Theorem \ref{thm:1D}), it is not necessary to renormalize the interaction $\bW$ and the weak convergence 
$$
\frac{k!}{T^k} \Gamma_{\lambda}^{(k)} \rightharpoonup \int_{\mathfrak{H}^{1-p}} |u^{\otimes k}\rangle \langle u^{\otimes k}| \,d\eta(u), \quad \forall k\ge 1
$$
is enough to imply the lower bound
$$\liminf T^{-2}\Tr[\bW \Gamma_\lambda] \ge \int \mathcal{D}[u]  \,d\eta(u)$$
by some sort of Fatou's lemma. In higher dimensions, such a compactness argument is not sufficient (since $\bW$ now depends on $T$) and we thus have to  develop quantitative estimates. 

We will localize the problem to finite dimensions (on low kinetic energy modes). Recall  that  for any projection $P$ on the one-body Hilbert space and for any quantum state $\Gamma$ on Fock space, we can construct the localized state $\Gamma_P$ on Fock space with density matrices 
\begin{equation}
(\Gamma_P)^{(k)}=P^{\otimes k}\Gamma^{(k)}P^{\otimes k},\qquad \forall k\geq1
\label{eq:GammaV-k}
\end{equation} 
(see  \cite{Lewin-11} for a general discussion  of the localization method on Fock space). In finite dimensions, a quantitative version of the de Finetti measure can be constructed explicitly  using the  \emph{lower symbol} (or {\em Husimi function}) 
\begin{equation} \label{eq:Husimi}
d\mu_{P,\Gamma}^{\eps}(u):=(\eps\pi)^{-\Tr(P)}\pscal{\xi(u/\sqrt{\eps}),\Gamma_P \xi(u/\sqrt{\eps})}_{\cF(P\gH)} d u
\end{equation}
where $du$ is the usual Lebesgue measure on $P\gH \simeq \C^{\Tr(P)}$ and $\xi(u)$ is the {\em coherent state}
$$
\xi(u):= \exp(a^\dagger(u)-a(u))  |0\rangle = e^{-\norm{u}^2/2} \bigoplus_{n \ge 0} \frac{1}{\sqrt{n!}} u^{\otimes n}. 
$$
The following theorem, taken from \cite[Lemma 6.2 and Remark 6.4]{LewNamRou-14d}, is an extension of a fundamental result of Christandl-K{\"o}nig-Mitchison-Renner \cite{ChrKonMitRen-07} in the canonical setting.  
\begin{theorem}[Quantitative quantum de Finetti theorem] For all $\eps>0$ and $k\in\N$,
 for $\Tr [P]=n$ and $\cN=\int a_x^* a_x$ the number operator on Fock space, 
\begin{align}
\Tr \left| k!\eps^k\Gamma^{(k)}_P-\int_{P\gH}|u^{\otimes k}\>\<u^{\otimes k}|\;d\mu^\eps_{P,\Gamma}(u) \right| \leq \eps^k \sum_{\ell=0}^{k-1}{k\choose \ell}^2  \frac{(k-\ell +n-1)!}{(n-1)!}\tr \left[ \cN^{\ell}\Gamma_P\right].
\label{eq:quantitative}
\end{align}
\end{theorem}
In applications, if $\langle \cN \rangle_\Gamma \sim N$ large, then we will choose $\eps=N^{-1}$ and obtain an error of order $O(n/N)$ for any fixed $k$.  

Finally, let us remark that the localization procedure behaves nicely with the relative entropy. The following Berezin-Lieb type inequality is taken from \cite[Theorem 7.1]{LewNamRou-14d}, whose proof goes back to the techniques in~\cite{Berezin-72,Lieb-73b,Simon-80}. 

\begin{theorem}[\textbf{Relative entropy: quantum to classical}] \label{thm:rel-entropy}\mbox{}\\
Let $\Gamma$ and $\Gamma'$ be two states on $\cF(\gH)$. Let $\mu_{P,\Gamma}^\eps$ and $\mu_{P,\Gamma'}^\eps$ be the lower symbols defined in \eqref{eq:Husimi}.  Then we have
\begin{equation}
\cH(\Gamma,\Gamma')\geq \cH(\Gamma_P,\Gamma'_P)\geq \cHcl(\mu_{P,\Gamma}^\eps,\mu_{P,\Gamma'}^\eps).
\label{eq:Berezin-Lieb}
\end{equation}
\end{theorem}

\subsection{Correlation estimates} We will eventually take the spectral projections  
$$ P = {\mathds 1}(h \leq K), \quad Q={\mathds 1}(h > K)$$
for some energy cut-off $K$. Here $K$ cannot be too large because we need ${\rm Tr} P \ll T$, to control the error in the quantitative quantum de Finetti theorem. On the other hand, we will take $K\to \infty$ simultaneously as $T\to \infty$ in order to control the localization error of the interaction energy. This task is the main challenge in the higher dimensional case.

We will rewrite the quantum interaction as
\begin{equation}\label{eq:discu renorm quant}
  \bW = \frac{1}{2}\sum_k \hat{w} (k) \left| \sum_p a^\dagger_{k+p}a_p  - \left\langle \sum_p a^\dagger_{k+p}a_p \right\rangle_{\Gamma_{0}} \right| ^2 
  \end{equation}
with the creation/annihilation operators of Fourier modes and perform the localization procedure on each mode. It turns out that the most problematic term comes from the zero-momentum mode, where we need to establish an estimate of the form 
\begin{align} \label{eq:corr-est-0}
\left\langle \Big( \mathcal{N}_Q -\langle \mathcal{N}_Q \rangle_{\Gamma_0} \Big)^2 \right\rangle_{\Gamma_\lambda}  =  o(T^2),  \quad \mathcal{N}_Q = \int_{\Omega} a_x^* Q_x a_x d x.
\end{align}

Note that in the eligible range of $K$, the expectation  $\langle \mathcal{N}_Q\rangle_{\Gamma_\lambda}$ grows much faster than $T$ and \eqref{eq:corr-est-0} relies on an  important cancelation by $\langle \mathcal{N}_Q\rangle_{\Gamma_0}$. In fact, in the homogeneous case considered here, the zero-momentum mode is the only one that gets renormalized (because of translation
invariance), so it is natural that it is the most troublesome.

The desired bound \eqref{eq:corr-est-0} is supported by the physical intuition that the free and interacting Gibbs states do not differ much in the high energy modes (since particles move too fast to interact).  The justification of \eqref{eq:corr-est-0} needs several new ideas that we will explain below.

\medskip

\noindent
{\bf Step 1.} First, we prove a weaker version of \eqref{eq:corr-est-0}:  
$$
|\langle \mathcal{N}_Q \rangle_{\Gamma_\lambda}-\langle \mathcal{N}_Q \rangle_{\Gamma_0}| = o(T).
$$

This can be done by a Feynman-Hellmann type argument.  More precisely, for any one-body self-adjoint operator $A<h$, by using Gibbs' variational principle for the free Gibbs state with $h$ replaced by $h-A$, we obtain 
$$
{\rm Tr} \Big( A\big(\Gamma_\lambda^{(1)}-\Gamma^{(1)}_0\big)\Big)\leq \mathcal{H}(\Gamma_\lambda,\Gamma_0)+{\rm Tr} \left(A\left(\frac{1}{e^{h-A}-1}-\frac{1}{e^{h}-1}\right)\right).
$$
Then by choosing $A=h^\alpha B h^\alpha$ for a general self-adjoint operator $B$ and a small constant $\alpha>0$, we deduce from a Klein-type inequality ~\cite[Proposition~3.16]{OhyPet-93} that 
$$
{\rm Tr}\left| h^\alpha \frac{\Gamma_\lambda ^{(1)} - \Gamma_0 ^{(1)}}{T} h ^\alpha \right| \leq C_\alpha,
$$
which in turn implies the desired bound. 

Unfortunately the Feynman-Hellmann principle does not work with two-body perturbations as we cannot afford to destroy the positivity of the interaction (for the one-body part this is not a problem). Hence, \eqref{eq:corr-est-0} is much harder to prove.

\medskip

\noindent
{\bf Step 2.}  From the previous step, we can rewrite  \eqref{eq:corr-est-0} as the true variance estimate as 
$$
\left\langle \Big( \mathcal{N}_Q -\langle \mathcal{N}_Q \rangle_{\Gamma_\lambda} \Big)^2 \right\rangle_{\Gamma_\lambda} =  o(T^2).
$$
The key idea is to reduce this two-body estimate to a one-body estimate. More precisely, we will approximate the variance by the {\em linear response} of $\mathcal{N}_Q$:
\begin{equation}\label{eq:intro correl}
T^{-2} \left\langle \left| \mathcal{N}_Q - \left\langle \mathcal{N}_Q  \right\rangle_{\Gamma_{\lambda}} \right| ^2 \right\rangle_{\Gamma_\lambda} = T^{-1} \partial_{\varepsilon} \left( \left\langle \mathcal{N}_Q  \right\rangle_{\Gamma_{\lambda,\varepsilon}} \right)_{|\varepsilon = 0} + o(1)
\end{equation}
where 
$$
 \Gamma_{\lambda,\varepsilon}:=\mathcal{Z}_{\lambda,\varepsilon}^{-1} \exp\left( -\frac{1}{T} \left( \mathbb{H}_\lambda - \varepsilon \mathcal{N}_Q \right)\right).
$$

Note that if a  self-adjoint operator $A$ commutes with $\mathbb{H}_\lambda$, then we have Kubo
formula for the canonical correlation \cite{Kubo-66}
$$
T ^{-1} \partial_\varepsilon  \left( \langle A \rangle_{\Gamma_{\lambda,\varepsilon}} \right)_{|\varepsilon = 0} = T^{-2}  \langle \left| A  - \langle A \rangle_{\Gamma_{\lambda}} \right| ^2 \rangle_{\Gamma_{\lambda}}. 
$$
However, $\mathcal{N}_Q$ does not commute with $\mathbb{H}_\lambda$ and it is useful to interpret the linear response as an averaged version of the quantum variance. We have the following abstract result \cite[Lemmas 7.3, 7.4]{LewNamRou-18}.

\begin{theorem}[Linear response and averaged quantum variance] Let $H,A$ be self-adjoint operators on a separable Hilbert space  such that $A$ is $H$-bounded and $\tr[e^{-sH}]<\ii$ for any $s>0$. Consider the quantum state 
$
\Gamma= e^{-H}/\Tr [e^{-H}]
$
and the \emph{quantum $s$-variance} 
$$
\Vars_\Gamma(A):=\tr\left[A\Gamma^s A\Gamma^{1-s}\right]-\big(\tr[A\Gamma]\big)^2.
$$
Then 
$$
\frac{d}{d\eps} \left[ \frac{\tr\left( Ae^{-H+\eps A}\right)}{\tr(e^{-H+\eps A})} \right]_{|\eps=0} =\int_0^1 \Vars_\Gamma(A) ds
$$
and
$$
 0\leq \Vars_\Gamma(A)\leq \Varq_\Gamma(A) \le \Vars_\Gamma (A)- \frac12 \tr \left( \left[A,H\right]^2 \Gamma \right),\quad \forall s\in [0,1].
$$
\end{theorem}

In applications, the difference between the linear response and the true quantum variance in \eqref{eq:intro correl} can be estimated in term of the commutator $[\cN_Q,\bH_\lambda]^2$. In general, if $\tr[h^{-p}]<\ii$ for some $1<p\leq 2$, then we can show that  
$$
- T^{-2}[\cN_Q, \bH_\lam]^2  \le C \Big(  T^{-2}\cN^4+ T^{2p-2}\cN^2 \Big)
$$
and consequently,
$$
- T^{-2} \Tr \left[  [\cN_Q, \bH_\lam]^2 \Gamma_\lambda \right]   \le C (T^{2p-3} + T^{2(2p-3)} ).
$$
This error can be controlled as soon as ${\rm Tr}[h^{-p}]<\infty$ for some $p<3/2$. This technical condition holds in 2D (we can choose $p>1$ arbitrarily) but barely fails in 3D.  

\medskip

\noindent
{\bf Step 3.}  Now it remains to estimate the linear response on the right side of \eqref{eq:intro correl}. In principle, the first derivative can be estimated using Taylor's expansion, namely
$$
g'(0) = \frac{g(\eps)-g(0)}{\eps} - \frac{\eps}{2} g''(\theta_\eps) \text{ for some }\theta_\eps \in (0,\eps).
$$

In applications, the difference $g(\eps)-g(0)$ can be bounded similarly to Step 1 (this involves only one-body terms). The second derivative involves the third moment 
$$\langle (\cN_Q-\langle \cN_Q \rangle_{\Gamma_\lambda})^3\rangle_{\Gamma_\lambda}$$
but fortunately a rough estimate of this term (without taking the cancellation into account) is sufficient for our analysis in 2D (this is another place where we lose a lot and could not treat the 3D case). Thus \eqref{eq:intro correl} holds, leading to the key correlation estimate \eqref{eq:corr-est-0}. 

\subsection{Relative free energy convergence} Now we are ready to relate the quantum problem \eqref{eq:rel-free} with the classical one \eqref{eq:rel-free-classical}. As explained before, we will restrict the quantum problem to finite dimensions using the projection $P=\1(h\le K)$ with $1\ll K \ll T$. The correlation estimate \eqref{eq:corr-est-0} and the quantitative quantum de Finetti estimate \eqref{eq:quantitative} imply that
\begin{align*}
T^{-2} {\rm Tr} [\bW \Gamma_\lambda^{(2)}]  &= T^{-2} {\rm Tr} [\bW P^{\otimes 2}\Gamma_\lambda^{(2)}P^{\otimes 2}]  + o(1)\\
&=  \frac{1}{2} \int_{P\gH} \cD_K [u] d\mu_{P,\lambda}(u) + o(1)
\end{align*}
where $\mathcal{D}_K$ is the truncated renormalized interaction in Lemma \ref{lem:re-interaction} and $\mu_{P,\lambda}$ is the lower symbol of $\Gamma_\lambda$ associated with the projection $P$ and the scale $\eps=T^{-1}$. Using this together with the Berezin-Lieb inequality \eqref{eq:Berezin-Lieb}, 
$$
\cH(\Gamma_\lambda,\Gamma_0) \ge \cH((\Gamma_\lambda)_P,(\Gamma_0)_P)    \ge \cHcl(\mu_{P,\lambda}, \mu_{P,0}),
$$
and using the classical variational principle \eqref{eq:rel-free-classical} we find that 
\begin{align} \label{eq:partition-lwb-0}
 -\log \frac{\cZ_\lambda}{\cZ_0} &= \cH(\Gamma_\lambda,\Gamma_0) + T^{-2} \Tr [\bW\Gamma_\lambda] \nn\\
&\ge \cHcl(\mu_{P,\lambda}, \mu_{P,0})+\frac12\int_{P\gH} \cD_K [u] d\mu_{P,\lambda}(u) + o(1)
 \nn\\
&\ge -\log \left( \int_{P\gH} e^{-\cD_K [u]} \; d\mu_{P,0}(u)\right) + o(1).
\end{align}

Moreover, the lower symbol $\mu_{P,0}$ of $\Gamma_0$ is very close to the cylindrical projection $\mu_{0,K}$ of the free Gibbs measure $\mu_0$ on $P\gH$. In fact, by using the Peierls-Bogoliubov inequality $\langle x, e^A x\rangle \ge e^{\langle x,A x\rangle}$ \cite[Theorem 2.12]{Carlen-12} we can show that  if $\Tr[h^{-p}]<\infty$, then 
\begin{equation}\label{eq:mu_L1}
\norm{\mu_{P,0}-\mu_{0,K}}_{L^1(P\gH)}\leq C T^{-1}K^{p+1}.
\end{equation}
Thus choosing $K$ appropriately we obtain 
\begin{align*}
-\log \frac{\mathcal{Z}_\lambda}{\mathcal{Z}_0}  \ge  -\log \left( \int_{P\gH} e^{-\cD_K [u]} \; d\mu_{0,K}(u)\right) + o(1) = -\log z_r + o(1).
\end{align*}
The matching upper bound can be obtained by a trial state argument, and we arrive at 
\begin{align*}
-\log \frac{\mathcal{Z}_\lambda}{\mathcal{Z}_0} = -\log z_r + o(1).
\end{align*}

\subsection{Convergence of density matrices} The convergence of reduced density matrices follows by carefully refining various estimates in the above proof of the relative free energy convergence. 

Since the free Gibbs state is factorized under the localization, i.e.  
$$
\Gamma_0 = (\Gamma_0)_P \otimes (\Gamma_0)_Q
$$
up to a unitary equivalence, we have the following simple but very useful identity 
$$
\cH(\Gamma_\lambda,\Gamma_0)= \cH(\Gamma_\lambda, (\Gamma_\lambda)_P\otimes (\Gamma_\lambda)_Q) + \cH((\Gamma_\lambda)_P,(\Gamma_0)_P) +  \cH((\Gamma_\lambda)_Q,(\Gamma_0)_Q) .
$$
Since we have used only $\cH((\Gamma_\lambda)_P,(\Gamma_0)_P)$ in the lower bound \eqref{eq:partition-lwb-0}, we can deduce that 
$$
\cH(\Gamma_\lambda, (\Gamma_\lambda)_P\otimes (\Gamma_\lambda)_Q) \to 0, \quad \cH((\Gamma_\lambda)_Q,(\Gamma_0)_Q)\to 0.
$$ 

Next, we will use the (quantum and classical) Pinsker inequalities,
$$\cH(A,B)\geq \frac12(\tr|A-B|)^2,\qquad \cH_{\rm cl}(\mu,\nu)\geq \frac12 \Big(|\mu-\nu|(\gH)\Big)^2.$$
By the quantum Pinsker inequality and the above estimates we obtain
\begin{align}\label{eq:state-bound}
\Tr \left| \Gamma_\lambda - (\Gamma_\lambda)_P\otimes (\Gamma_\lambda)_Q \right|  \to 0, \quad \Tr \left| (\Gamma_\lambda)_Q - (\Gamma_0)_Q\right| \to 0.
\end{align}
In general, the bounds on states can be transferred to estimates on density matrices using the Cauchy-Schwarz type inequality 
$$
\Tr |\Gamma^{(k)} -\Gamma'^{(k)} | \le \left( \Tr |\Gamma-\Gamma'| \right)^{1/q'} \left( \Tr[ \cN^{qk}(\Gamma+\Gamma')]\right)^{1/q}, \quad \frac{1}{q}+\frac{1}{q'}=1. 
$$
Given that the error estimate in \eqref{eq:state-bound} is good enough to compensate for the divergence of the particle number expectation, we can deduce that, for the case of one-body density matrices, 
\begin{align} \label{eq:DM-PP-QQ}
\Tr\left| \Gamma^{(1)} -  P\Gamma_\lambda^{(1)} P - Q \Gamma_\lambda^{(1)}Q \right| \to 0, \quad  \Tr\left| Q\Gamma_\lambda^{(1)} Q -  Q\Gamma_0^{(1)} Q \right| \to 0.
\end{align}

Let us take a  closer look at $P\Gamma_\lambda^{(1)} P$. From the lower bound \eqref{eq:partition-lwb-0} and the classical Pinsker inequality 
we have 
$$
\norm{\mu_{P,\lambda} - \frac{e^{-\cD_K[u]} \; d\mu_{0,K}(u)}{\int_{P\gH} e^{-\cD_K[v]} \; d\mu_{0,K}(v) }}_{L^1(P\gH)} \to 0.   
$$
The quantum de Finetti estimate \eqref{eq:quantitative} then tells us that 
\begin{align} \label{eq:DM-PP}
\frac{1}{T}\Tr \left|  P\Gamma_\lambda^{(1)} P-  P \left( \int_{P\gH}|u\>\<u|\;d\mu (u) \right) P \right|  \to 0.
\end{align}
Putting \eqref{eq:DM-PP-QQ} and \eqref{eq:DM-PP} together, we conclude that 
$$
\Tr \left| \frac{1}{T} \left( \Gamma_\lambda^{(1)} -  \Gamma_0^{(1)} \right) - \int |u \rangle \langle u | \Big( d\mu(u) - d\mu_0(u) \Big) \right| \to 0
$$
by the triangle inequality. Let us explain some details for the one-body density matrices.

For the two-body density matrix, using the above strategy we can write
\begin{align} \label{eq:2-body-DM}
T^{-2}(\Gamma_\lambda ^{(2)} - \Gamma_0^{(2)} ) &=T^{-2} P^{\otimes 2} (\Gamma_\lambda ^{(2)}-\Gamma_0^{(2)}) P^{\otimes 2} + T^{-2} Q^{\otimes 2} (\Gamma_\lambda ^{(2)}-\Gamma_0^{(2)}) Q^{\otimes 2} \nn\\
& + T^{-2} P\Gamma_{\lambda} ^{(1)} P \otimes Q (\Gamma_\lambda ^{(1)}-\Gamma_0^{(1)}) Q 
+ T^{-2} Q (\Gamma_\lambda ^{(1)}-\Gamma_0^{(1)}) Q  \otimes P\Gamma_{\lambda} ^{(1)} P \nn\\
& + T^{-2} P(\Gamma_{\lambda} ^{(1)}-\Gamma_0^{(1)}) P \otimes Q \Gamma_0^{(1)} Q 
+ T^{-2} Q \Gamma_0 ^{(1)} Q \otimes P(\Gamma_{\lambda} ^{(1)}-\Gamma_0^{(1)}) P.
\end{align}
By following the above analysis, we can show that the first four terms on the right side of \eqref{eq:2-body-DM} converge in trace class. However, the last two terms are unbounded in trace norm, and hence $T^{-2}(\Gamma_\lambda ^{(2)} - \Gamma_0^{(2)})$ is also unbounded in trace norm. Nevertheless, the last two terms in \eqref{eq:2-body-DM} converge to $0$ in the Schatten norm $\gS^p$ with $p>1$, leading to  the desired convergence of $T^{-2}\Gamma_\lambda^{(2)}$ in  $\gS^p$. The same proof works for the convergence of higher density matrices. 

\subsection*{Acknowledgements.} This project has received funding from the European Research Council (ERC) under the European Union's Horizon 2020 Research and Innovation Programme (Grant agreements MDFT No 725528 and CORFRONMAT No 758620).

\end{document}